\documentclass[]{spie}  

\usepackage{amsmath,amsfonts,amssymb}
\usepackage[utf8]{inputenc}
\usepackage{graphicx}
\usepackage[colorlinks=true, allcolors=blue]{hyperref}
\usepackage[perpage]{footmisc}

\usepackage{tabularx,adjustbox,booktabs}

\newcommand\clearrow{\global\let\rowmac\relax}
\clearrow

\newcolumntype{$}{>{\global\let\currentrowstyle\relax}}
\newcolumntype{^}{>{\currentrowstyle}}

\title{The NIRSpec Micro-Shutter Array: Operability and Operations After Two Years of JWST Science}

\author[a]{Katie Bechtold}
\author[b]{Torsten B\"oker}
\author[c]{David E. Franz} 
\author[b]{Maurice te Plate}
\author[d]{Timothy D. Rawle}
\author[a]{Rai Wu}
\author[e]{Peter Zeidler}

\affil[a]{Space Telescope Science Institute (STScI), Baltimore, MD 21218, USA}
\affil[b]{European Space Agency (ESA), ESA Office, STScI, Baltimore, MD 21218, USA}
\affil[c]{NASA Goddard Space Flight Center, Greenbelt, MD 20771, USA}
\affil[d]{European Space Agency (ESA), ESAC, Madrid, Spain}
\affil[e]{AURA for the European Space Agency, ESA Office, STScI, Baltimore, MD 21218, USA}

\authorinfo{Send correspondence to K.B.  E-mail: kbechtold@stsci.edu}

\begin{document} 
\maketitle

\begin{abstract}
The Near Infrared Spectrograph (NIRSpec) on the James Webb Space Telescope affords the astronomical community an unprecedented space-based Multi-Object Spectroscopy (MOS) capability through the use of a programmable array of micro-electro-mechanical shutters. Launched in December 2021 and commissioned along with a suite of other observatory instruments throughout the first half of 2022, NIRSpec has been carrying out scientific observations since the completion of commissioning. These observations would not be possible without a rigorous program of engineering operations to actively monitor and maintain NIRSpec's hardware health and safety and enhance instrument efficiency and performance. Although MOS is only one of the observing modes available to users, the complexity and uniqueness of the Micro-Shutter Assembly (MSA) that enables it has presented a variety of engineering challenges, including the appearance of electrical shorts that produce contaminating glow in exposures. Despite these challenges, the NIRSpec Multi-Object Spectrograph continues to perform robustly with no discernible degradation or significant reduction in capability.

This paper provides an overview of the NIRSpec micro-shutter subsystem's state of health and operability and presents some of the developments that have taken place in its operation since the completion of instrument commissioning.
\end{abstract}


\section{INTRODUCTION}
\label{sec:intro}

A key mission goal of the James Webb Space Telescope (JWST) is the study of galaxy formation in the early universe, which involves acquiring spectra of a large number of high-redshift galaxies. However, the sparsity of such targets on the sky presents a challenge to the efficiency of any potential surveys. In order to enable a statistically meaningful catalog of high-redshift spectra, it is essential to ``multiplex'' such observations, i.e. to obtain spectra of many faint galaxies simultaneously. This ability is provided by the multi-object spectroscopy (MOS) mode \cite{ferruit22} of NIRSpec, which has quickly become one of the most scientifically rich observing modes of JWST. 

After the launch of JWST on 25 December 2021, the NIRSpec health and performance in orbit was verified as part of the overall JWST commissioning campaign. The numerous NIRSpec commissioning activities \cite{2022SPIE12180E..0WB} were planned, executed, and analyzed by ESA and AURA instrument scientists, supported by engineers and subject matter experts from Airbus and NASA as well as flight systems engineers at the Space Telescope Science Institute (STScI). Following the JWST commissioning phase, NIRSpec MOS was declared ready for science operations on 1 July 2022. Science and mission operations are conducted from STScI in Baltimore, Maryland, with the support of subject matter experts around the globe. 

Two years into JWST science operations, it has become abundantly clear that the observatory in general, and the NIRSpec instrument in particular, are indeed performing at or above pre-launch expectations. This is demonstrated by the slew of JWST science results, with more than 500 refereed publications, numerous press releases, and extensive media coverage since June 2022. Many of these results were obtained using NIRSpec data. In particular, studies of the distant universe have been revolutionized by the availability of the NIRSpec MOS mode, because of its ability to obtain high-quality spectra of dozens of faint galaxies in a single exposure.

\section{BACKGROUND}
\label{sec:background}

\subsection{Instrument overview}
\label{sec:instrument-overview}

NIRSpec is one of four science instruments inside the Integrated Science Instrument Module (ISIM) onboard JWST, the others being the Near-Infrared Camera (NIRCam)\cite{NIRCam}, Near-Infrared Imager and Slitless Spectrograph (NIRISS)\cite{FGS+NIRISS}, and Mid-Infrared Instrument (MIRI)\cite{MIRI}.

Sensitive to light within the wavelength range of 0.6–5.3 $\mu$m, NIRSpec offers observers four science observing modes: multi-object spectroscopy (MOS), integral field spectroscopy (IFS), fixed-slit spectroscopy (FSS), and bright object time series (BOTS) spectroscopy. The subject of this manuscript, NIRSpec's Micro-shutter Assembly (MSA), is used in performing MOS observations and their associated MSA target acquisition (MSATA) activities, as well as undispersed imaging observations to verify target acquisition and pointing accuracy.

\begin{figure} [htbp]
\begin{center}
\includegraphics[width=0.6\linewidth]{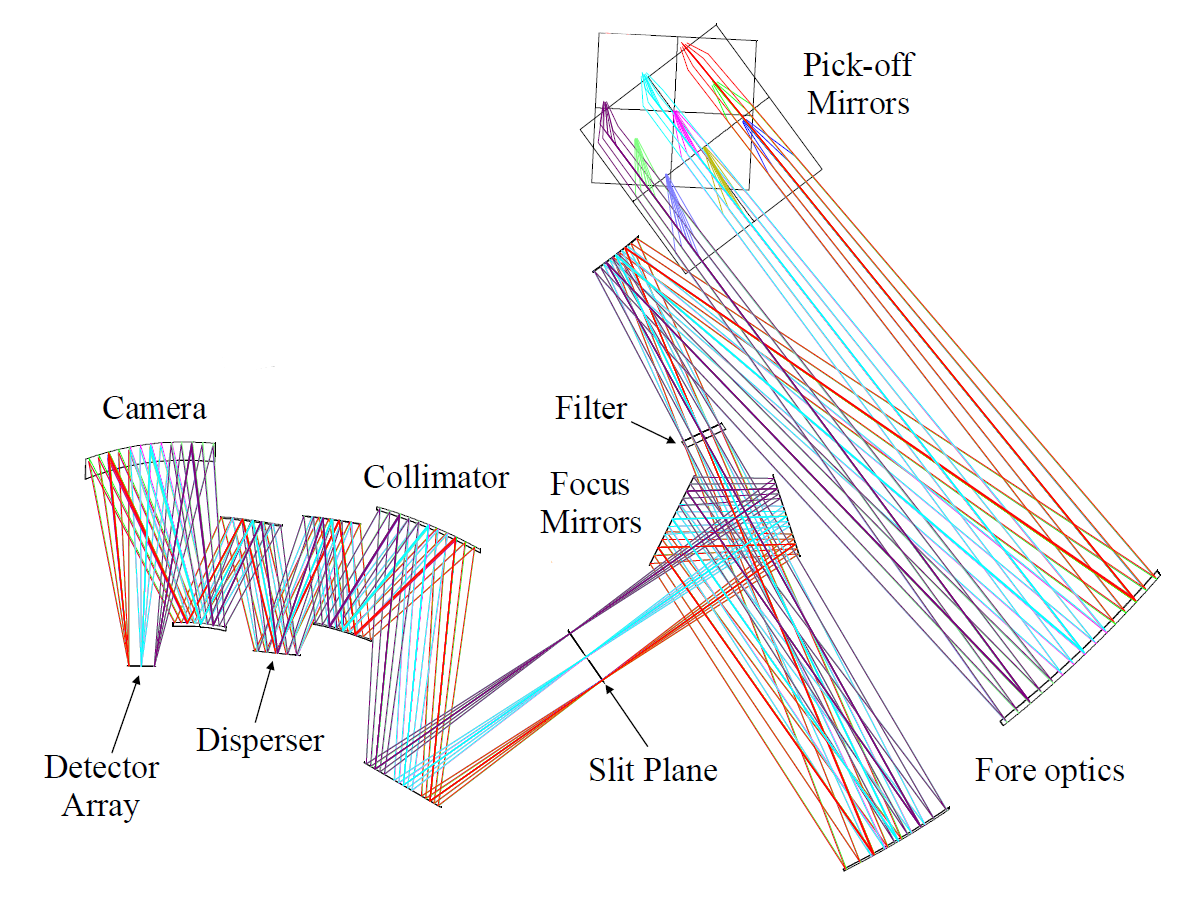}
\end{center}
\caption[lightpath] 
{ \label{fig:lightpath} 
NIRSpec light path for external observations. \cite{jakobsen22}}
\end{figure}

\begin{figure} [htbp]
\begin{center}
\includegraphics[width=0.7\linewidth]{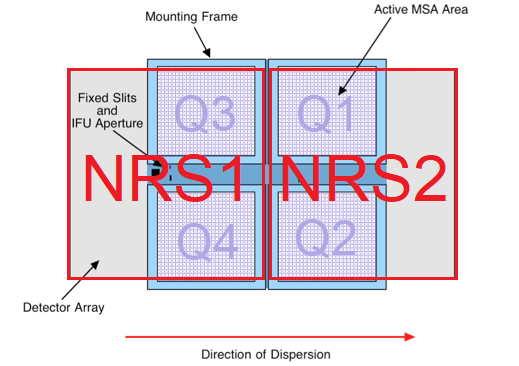}
\end{center}
\caption{MSA orientation in relation to the two-detector array. Light passing through the micro-shutters is dispersed onto the NIRSpec detectors, labeled NRS1 and NRS2.}
{ \label{fig:ds-orientation}}
\end{figure}

NIRSpec's optical path is a largely reflective system of fourteen mirrors (see Figure \ref{fig:lightpath}) \cite{jakobsen22}. A filter wheel assembly containing six interchangeable filters in the fore optics precedes the slit plane containing the MSA. In the light path after the slit plane, the system's collimator and camera optics are coupled by a grating wheel assembly containing seven interchangeable dispersive elements. After being dispersed or reflected, light is focused onto the focal plane assembly, which contains two abutting sensor chip assemblies. Figure \ref{fig:ds-orientation} is a schematic view of the MSA layout, projected onto the detector plane.

\subsection{JWST science operations paradigm}
\label{sec:science-execution}

Observations are autonomously executed on JWST in an event-driven operations paradigm\cite{elliottOSS} in which observations are sequenced in units of ``visits'' collectively comprising a weekly Observation Plan (OP), with commanding of instruments in general performed by the Operations Scripts Subsystem (OSS). In addition to external exposures, visits can encompass internal calibration and engineering activities, including some specific to the NIRSpec MSA as exemplified in Sections \ref{sec:esd} and \ref{sec:osd}. All visits are specified by observers or science instrument support staff using the Astronomer's Proposal Tools\footnote{\url{https://www.stsci.edu/scientific-community/software/astronomers-proposal-tool-apt}} (APT) software interface.

Instrument commanding can also be performed via real-time communication from the mission operations center when in contact via NASA's Deep Space Network (DSN). However, whereas the flight operations team enjoyed nearly around-the-clock DSN communication with the observatory in the commissioning phase of the mission, typical DSN coverage in normal operations is around eleven hours per day, much of it outside regular business hours. In addition, as a matter of policy, any anticipated nontrivial real-time commanding in normal operations—in contrast to commissioning—is scheduled to take place in dedicated time frames when the observatory is not simultaneously performing science observations. These new constraints on real-time commanding have spurred the evolution of some previously real-time engineering activities to be performed as part of an OP, a trend that is reflected in some of the developments presented in this paper.

\section{MICROSHUTTER ARRAY FUNCTIONALITY}
\label{sec:msa-functionality}

The MSA occupies NIRSpec's slit plane with a cruciform structure supporting a 2$\times$2 mosaic of micro-shutter quadrants. Also embedded in this structure are five permanently open fixed slits (FS) and an integral field unit (IFU) aperture that is only opened to the light path when IFU observations are taking place.

\begin{figure} [htbp]
\begin{center}
\begin{tabular}{c}
\includegraphics[height=4cm]{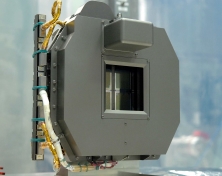}
\includegraphics[height=4cm]{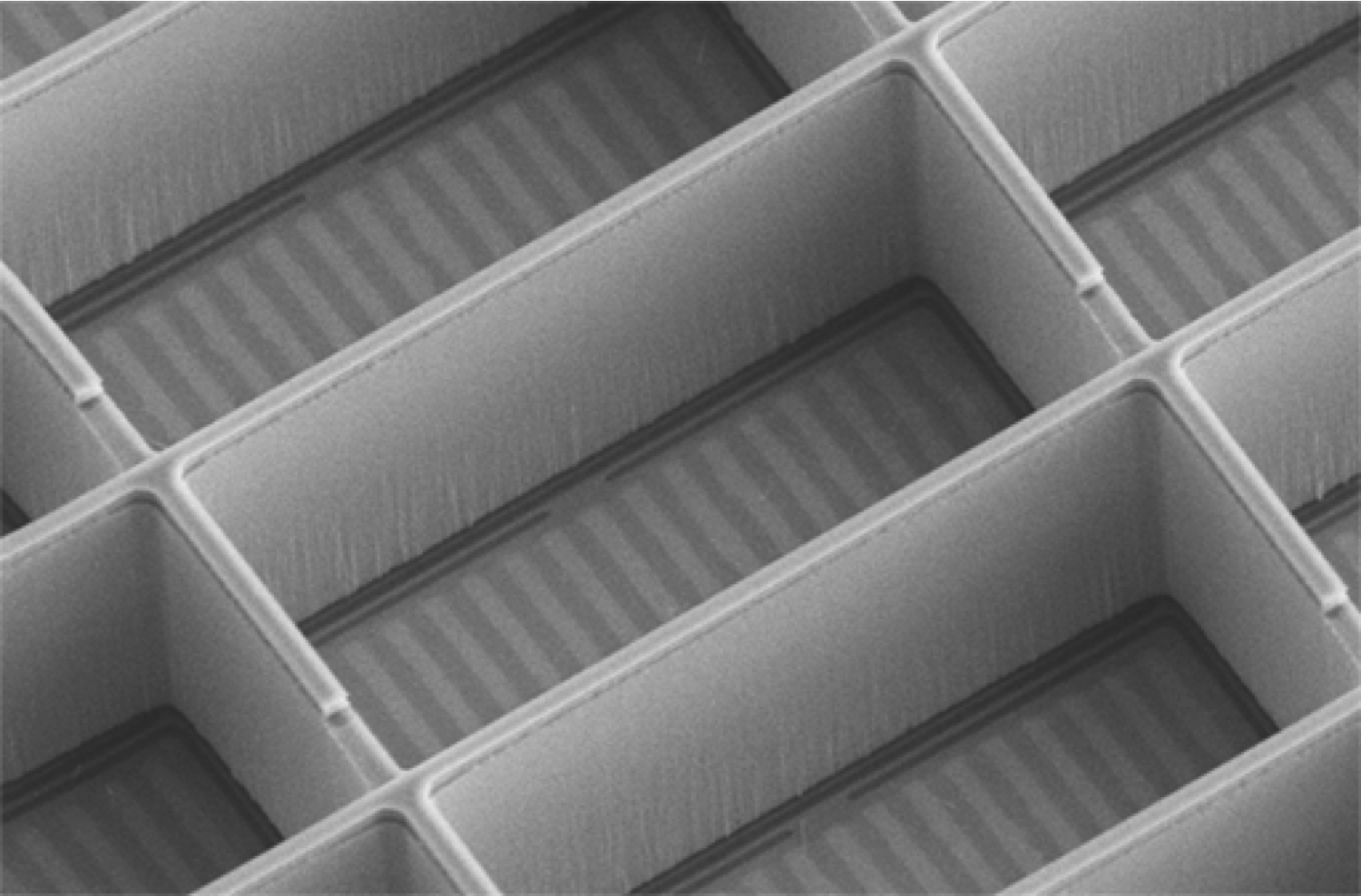}
\end{tabular}
\end{center}
\caption{Left: Micro shutter array framed by assembly with cover and translation stage plate. Right:  Microscopic view of shutters shown within the grid of ``crates.'' \textit{Credit: NASA}}
{ \label{fig:msaphotos}}
\end{figure}

Each micro-shutter quadrant is an array of 62,415 shutters etched from a semiconductor wafer: 171 lines of shutters in the spatial direction by 365 lines of shutters in the cross-dispersion direction, with each shutter measuring approximately 78$\times$178$\mu$m. Figure \ref{fig:msaphotos} includes a view of the micro-shutter array framed by the full assembly and a microscopic view of individual micro-shutters. Each shutter opens by rotating 90° on a hinge within a grid structure of walls parallel to the light path. Integral torsion bar springs keep the shutters closed when they are not being actively addressed. Each shutter has an electrode embedded in the shutter door that is connected on the front of the array to the side rails (171 side) and another electrode on the grid walls that is connected on the back of the array to the power supply (365 side). 

The MSA is controlled by two redundant sets of Micro-shutter Control Electronics (MCE) that share a common interface to the MSA itself. One set (MCE B) serves as a cold spare; thus far in flight only the primary set (MCE A) has been needed. Together with the instrument flight software, these elements make up the Micro-shutter Subsystem (MSS).

\begin{figure} [htbp]
\begin{center}
\includegraphics[width=\linewidth]{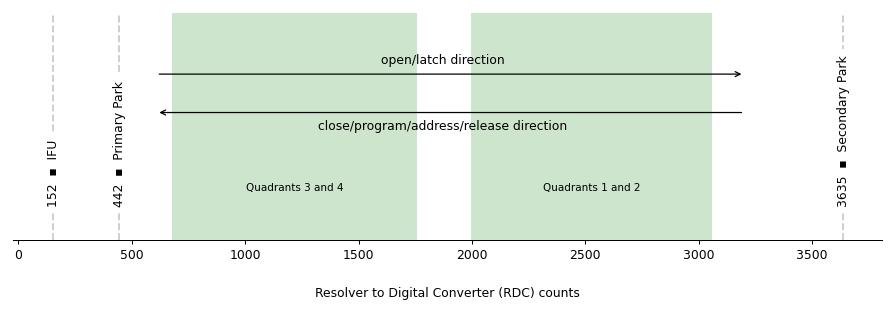}
\end{center}
\caption{Magnet arm resting positions (MCE A side) and movement directions. Not portrayed are two additional positions: Launch Lock, which was used for securing the magnet arm for JWST launch but is not used in normal operations, and Primary Park Overshoot (PPO), which was added after commissioning as described in Section \ref{sec:end-of-ifu-visit-hold-mask}.}
{ \label{fig:magarm-positions}}
\end{figure}

The process of programming a shutter pattern involves magnetic actuation in concert with electrostatic latching. The bottom of each shutter door is coated with strips of magnetic material so that as a quadrupole magnet mounted on a motor-driven ``magnet arm'' passes across the array moving from the Primary Park position to the Secondary Park position (indicated in Figure \ref{fig:magarm-positions}), all (mechanically operable) shutters are opened flush with their hinge-side walls. Opposite potentials (``open/latch voltage'') on 171-side and 365-side electrodes create an electrostatic force that latches the shutters in their open position. As the magnet arm then completes the cycle by moving across the array in the opposite direction and returning to the Primary Park position, the difference in potential between the wall and door electrodes on shutters that are to be closed is momentarily dropped, dissipating the electrostatic charge for long enough that the force of the torsion bar springs causes them to close. Conversely, a potential difference (``close/program/address voltage'') is maintained for shutters that are to remain open. These array potentials are synchronously changed as the magnet moves in the programming direction to apply a desired pattern on the shutter array.

At the end of the programming cycle, a low potential difference (``hold voltage'') is used to maintain the state of the shutters for use in observations. The full cycle to program a shutter pattern takes approximately two minutes including time for MCE commanding to load a shutter pattern bitmap, sweep the magnet arm over the array, and perform post-move application of a ``hold mask'' to mitigate glow caused by various types of leakage currents; see Section \ref{sec:resolved-issues} for a description of issues relating to this last step that have arisen and been resolved in normal operations. Each visit that involves programming the shutters concludes with a final magnet arm cycle to close all shutters over periods when they are not in use.

\section{OPERABILITY FACTORS}
\label{sec:operability-factors}

When specifying shutter patterns for their MOS observations, observers generally design a configuration of tens to hundreds of ``slitlets,'' small groups of shutters that are contiguous in the spatial dimension. Each slitlet contains a target \cite{2014SPIE.9149E..1ZK} using a nodding technique to allow for background subtraction. To plan observations, the observer needs to understand which shutters are operable such that they can be opened to the light path. In addition, most MOS observers—and a few FSS and IFS observers—opt to ensure targets are centered in the designated aperture(s) by preceding their science exposures with MSA Target Acquisition (MSATA), a method requiring users to select a set of reference stars subject to the same shutter operability conditions. The following sections describe factors that can render a shutter inoperable; note that a single shutter may be inoperable due to a combination of different factors.

\subsection{Geometric operability}
\label{sec:geometric-operability}

By design, the field stop framing the light path in the NIRSpec fore optics is ``undersized'' relative to the MSA apertures' collective field of view. As a result, a set of shutters around the edge of the MSA quadrants as a whole is vignetted, i.e. not illuminated by external sources. On the other hand, the optical path of NIRSpec's calibration lamps does not include the field stop, thus internal exposures illuminate the entire arrays. This set of vignetted shutters rendered inoperable by the geometry of the light path was definitively identified in commissioning and is expected to remain static throughout the mission.

\subsection{Mechanical operability}
\label{sec:mechanical-operability}

Warped or jammed shutters may become stuck in an open or closed position, temporarily or indefinitely. Shutters that are stuck in a closed position, classified as Failed Closed (FC), are less optically problematic than shutters that are stuck open, as sources that incidentally align with these Failed Open (FO) shutters may produce contaminating spectra. In the course of MSA ground testing, FO shutters identified before integration of the array with the rest of the NIRSpec optical assembly were permanently plugged, rendering them effectively FC.

To optimize planning of MSA observations using the MSA Planning Tool (MPT)\footnote{\url{https://jwst-docs.stsci.edu/jwst-near-infrared-spectrograph/nirspec-apt-templates/nirspec-multi-object-spectroscopy-apt-template/nirspec-mpt-plans\#gsc.tab=0}} in APT it is crucial for the user to know the operability status of the MSA, namely which micro-shutters can be commanded to open or close. The MPT interface highlights 171-side lines in which spectra could potentially be contaminated by light from unwanted sources being transmitted through FO shutters. For pattern planning purposes, shutters that are inoperable in any other respect—FC, short masked, or vignetted—are simply considered ``closed.''

\subsubsection{Monthly operability monitoring}
\label{sec:monthly-operability-monitoring}

To monitor the status of the MSA and, if needed, to update the shutter operability map, which is available in the JWST Calibration Reference Data System\footnote{\url{https://jwst-crds.stsci.edu/}} and the latest Project Reference Database used by APT and MPT, a NIRSpec calibration program\footnote{Cycle 1: Program ID 1488, Cycle 2: Program ID 4461, Cycle 3: Program ID 6637} is executed monthly. Visits in this program are scheduled and used to track the number of FC and FO shutters. Each month, four short exposures are obtained using an internal calibration lamp with an all-shutters-closed configuration, an all-shutters-open configuration, and two opposing ``$1\times1$ checkerboard'' configurations. The all-shutters-closed configuration is used to identify the FO shutters, while the all-shutters-open and checkerboard configurations are used to identify the lower and upper bounds of the FC population, respectively. In checkerboard patterns, half of the shutters are opened, alternating open and closed so that each closed shutter is bordered only by open shutters and vice versa; with these two distinct configurations, each operable micro-shutter is commanded, and the unique light pattern allows failed shutters to be easily identified.

\subsubsection{Updating the operability map}
\label{sec:updating-the-msop}

While updating the MSA operability map on a monthly basis is theoretically possible, based on the statistical failure rate of micro-shutters and the relatively work-intensive process, we have concluded that an operability map update rate of at most every six months is sufficient. That said, in two situations a more immediate reaction is necessary due to potential science exposure contamination: the appearance of an electrical short (see Section ~\ref{sec:electro-optical-operability-status}) and the appearance of a new FO shutter. While electrical and optical shorts contaminate exposures due to their strong thermal glow, FO shutters can contaminate other traces on the MSA as well as IFU exposures. Keeping the FO map as up-to-date as possible is important for MPT planning purposes and for pixel flagging in the IFU data reduction steps. Because we see that sometimes shutters are stuck only temporarily, a shutter is classified as FO or FC only if it was detected as such in three consecutive months.

\subsection{Electro-optical operability}
\label{sec:electro-optical-operability}

As a micro-electro-mechanical system (MEMS) device, the MSA is subject to occasional electrical shorts, in which an undesirable path for the flow of current is created due to factors like particulate contamination and flaws in the microscopic circuitry. Shorts can occur between two neighboring 171-side or 365-side lines (termed a ``Nearest Neighbor'' short) or between a 171-side and 365-side line (termed a ``Front-to-Back'' short). Shorts can also exist on the substrate itself.

\begin{figure} [htbp]
\begin{center}
\includegraphics[width=0.7\linewidth]{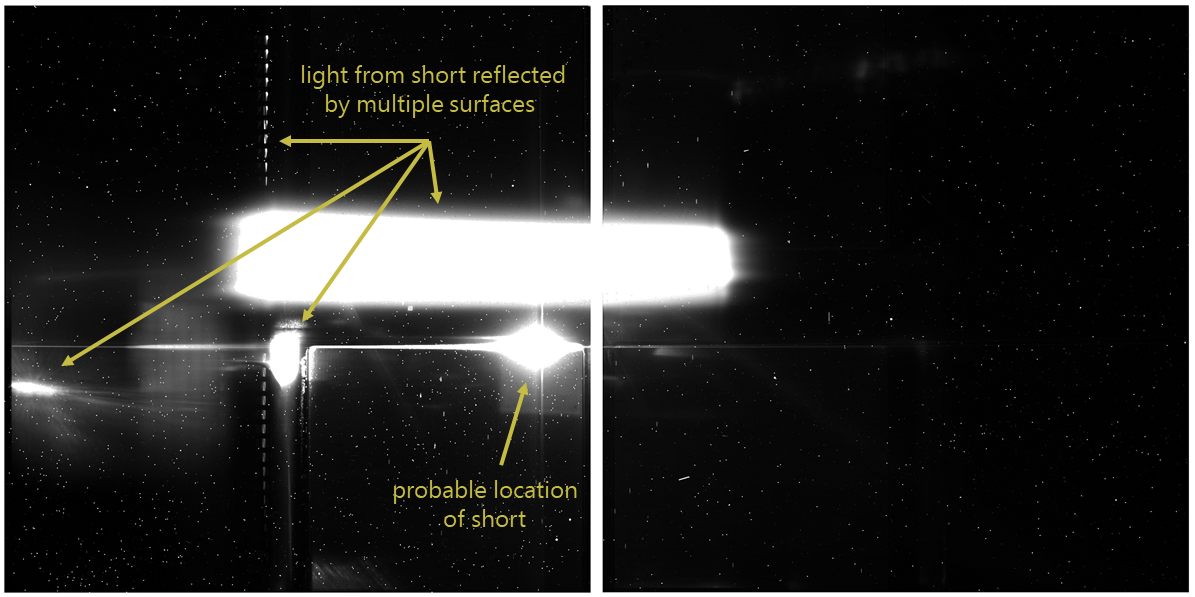}
\end{center}
\caption{Example of a target acquisition confirmation image with severe contamination from a short}
{ \label{fig:short_example} }
\end{figure}

\begin{figure} [htbp]
\begin{center}
\begin{tabular}{c}
\includegraphics[width=0.7\linewidth]{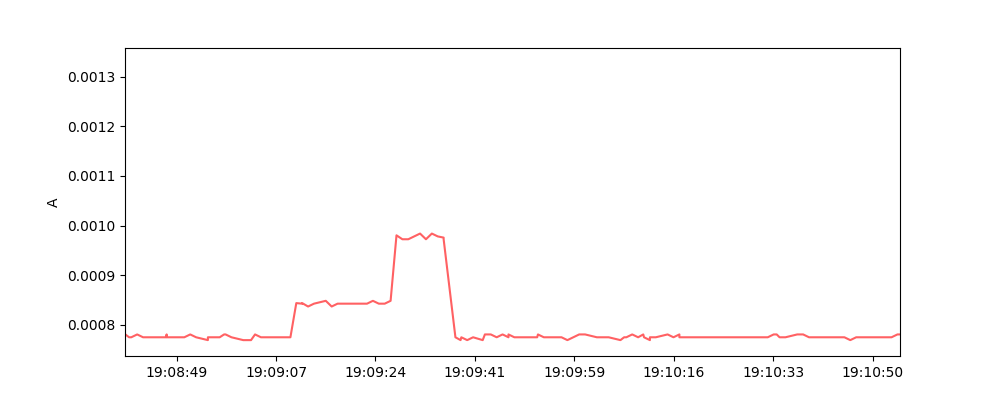} \\
\includegraphics[width=0.7\linewidth]{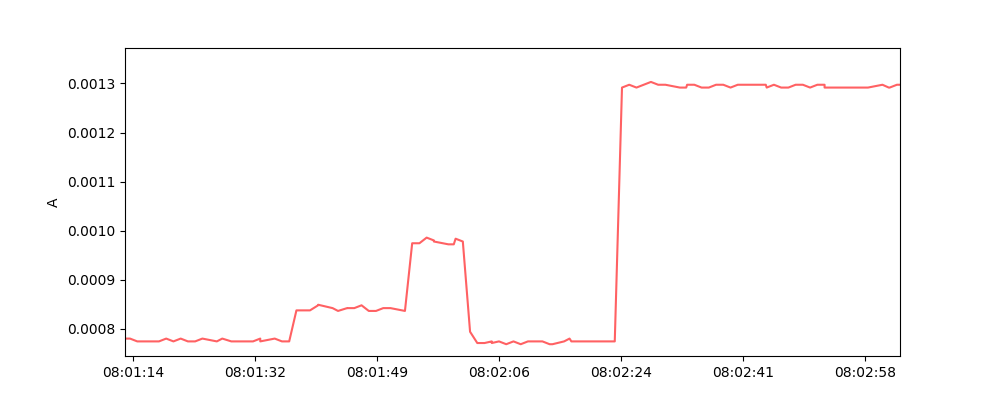}
\end{tabular}
\end{center}
\caption{Quadrant 4 365-side Vpp current in the course of a shutter reconfigurations without (top) and with (bottom) a short present.}
{ \label{fig:quad_current_plots} }
\end{figure}

Most shorts produce glow strong enough to significantly contaminate MOS exposures, as illustrated in Figure \ref{fig:short_example}, rendering them useless for scientific investigation.\footnote{This contamination could also impact IFS exposures (prior to a flight software update in June 2023; see Section \ref{sec:end-of-ifu-visit-hold-mask}), pointing verification imagery, and target acquisition confirmation imagery.} Most are also accompanied by elevated 365-side Vpp currents easily recognizable in engineering telemetry that is monitored on a daily basis as illustrated in Figure \ref{fig:quad_current_plots}. In integration and testing prior to launch, some shorts produced currents of about 1.2 mA above normal levels, high enough to generate not just a bright unwanted glow but also significant heat, raising the quadrant temperature and exposing the array to potential damage. By contrast, current levels seen in post-launch shorts typically involve elevations of around 0.4 mA above normal levels. It should be noted that after the shutters are closed at the end of a visit, application of the hold mask deactivates any short in the sense that the corresponding quadrant current returns to normal levels and the glow dissipates, but the anomalous current pathway can persist and can be re-activated when subsequent shutter patterns using the offending shutters are programmed.

While there is no hardware risk associated with these post-launch shorts, they are detrimental to observatory efficiency (because any unusable exposures will likely need to be repeated) and to MSA operability, as the masking of a shorts renders an entire row or column of shutters inoperable.

\subsubsection{Electrical shorts detection}
\label{sec:esd}

Electrical Short Detection (ESD) is an algorithmic process executed by the OSS to locate shorts by sequentially varying MCE line voltages to successively smaller subsets of shutter rows and columns and comparing the resulting quadrant currents to user-defined thresholds. For each set of shutter rows or columns (generically termed ``lines''), if the resulting current exceeds the threshold, the shutter set is considered to contain at least one line with a short. In this case smaller subsets of that shutter set are tested down to the individual row or column level until one or more lines with shorts are identified or all candidate lines have been tested without identifying individual lines with shorts. This algorithm was updated in May 2023 to better detect ``Nearest Neighbor'' shorts that the original algorithm often missed. Any individual lines found to have shorts are automatically added to a shorts mask maintained onboard in the MSS, and instrument subsystem engineers coordinate the configuration control processes to incorporate the updated mask into the formal project reference database system.

When a new short appears, ESD is scheduled as soon as practically possible. In commissioning, ESD was often initiated by real-time commanding, but for reasons described in Section \ref{sec:background}, in normal operations execution of such an activity as part of the OP is far preferred. Logistically, this involves creation of a visit using the ``NIRSpec MSA Short Detection'' APT template, specifying target quadrant(s) and thresholds, typically accompanied by an internal visit with exposures designed to surface the short in order to confirm its persistence and/or the success of any masking performed by the ESD process.

\subsubsection{Optical shorts detection}
\label{sec:osd}

\begin{figure} [htbp]
\begin{center}
\includegraphics[width=0.6\linewidth]{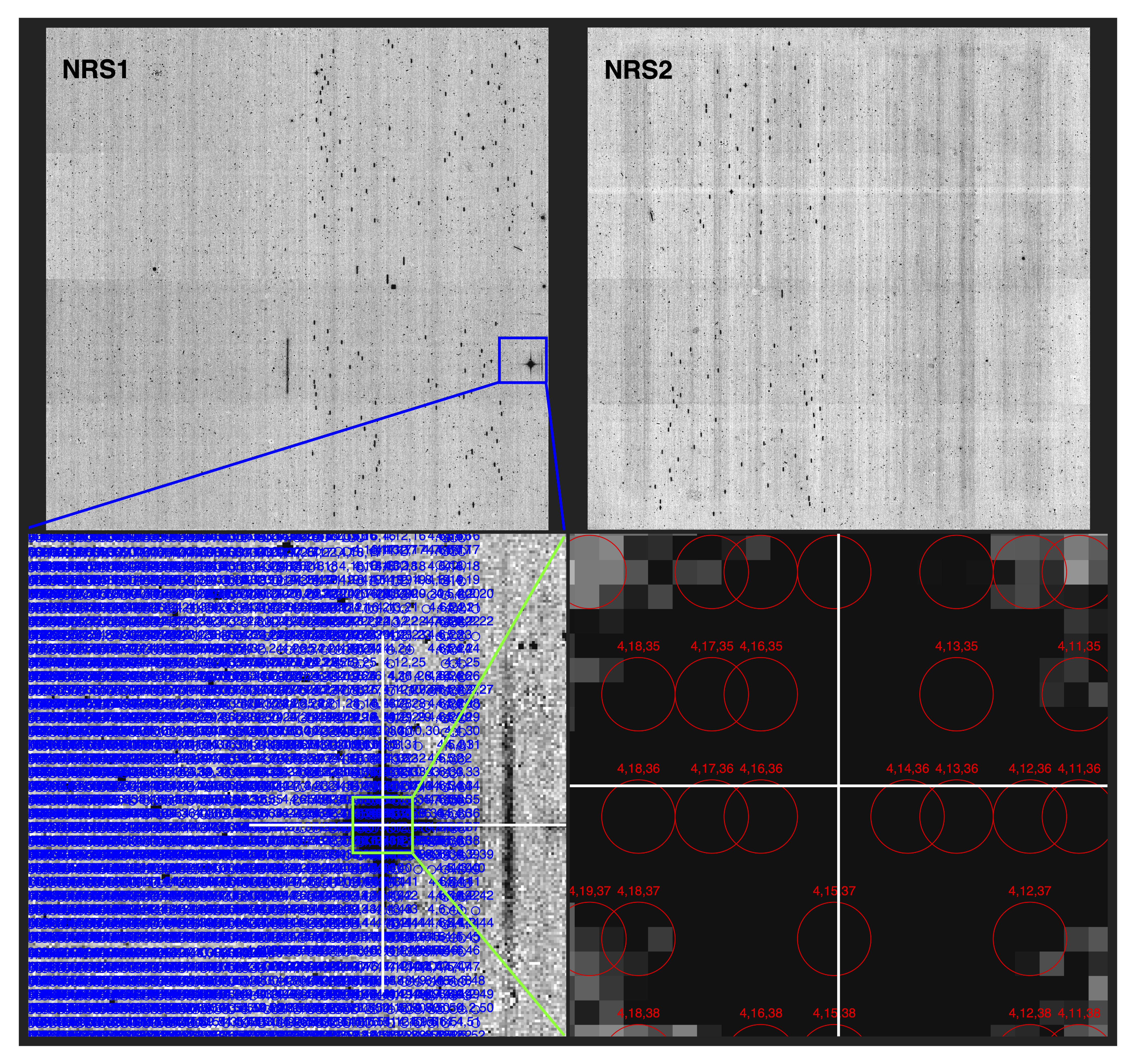}
\end{center}
\caption{Optical Shorts Detection. Upper row: Step 1. An undispersed exposure (Program ID 1213, Observation 1) showing a faint short on the right edge of Q4 as projected onto detector NRS1, with colors inverted for easier analysis. Three reflections of the glow are visible on the right, left and upper edge of the MSA Q4 mounting frame. Lower left: Step 2. Using the MSA–detector coordinate transformation derived in commissioning, a shutter map (blue) is superimposed on the image. Cross-hairs centering on the glow (white) are manually superimposed as an estimate of the short's location. Lower right: Step 3. The image is zoomed in to identify the shutter locations closest to the estimated center of the glow. Rows and columns adjacent to the estimated center of the glow are candidates used to create the test masks for Optical Shorts Detection.}
{ \label{fig:osd-process} }
\end{figure}

Because some shorts do not elevate quadrant currents to distinguishably abnormal levels in engineering telemetry, not all shorts can be located and automatically masked by the ESD process described in Section \ref{sec:esd}. In these cases, to locate the short we must employ Optical Shorts Detection (OSD), a manual process of analyzing exposures that show glow from a short to identify the rows and/or columns most likely to geometrically align with the short (see Figure \ref{fig:osd-process}) and using them as a basis in creating ``test'' short masks. A series of visits interleaving the application of the various test masks with exposures designed to induce the short is defined and added to an Observation Plan. After these visits execute, the exact location of the short can be deduced from analysis of the resulting exposures, identifying the test mask(s) that succeeded in preventing glow from the short. Only at this point can the process of updating the short mask onboard and incorporating it into the configuration-managed system commence.

Throughout commissioning and the first year of normal operations, the only means of applying short masks onboard in order to test them required performing ground-based real-time commanding in a complex series of hand-offs of instrument commanding control between ground-based controllers and OSS. As mentioned in Section \ref{sec:background}, real-time commanding is relatively disruptive in normal operations, and activities requiring many transfers of control between controllers and OSS especially so. This unwieldiness made it highly desirable to develop the means of performing OSD without the need for real-time commanding.

The ability to apply short masks onboard via simple, concise visits in an OP was developed after commissioning and deployed in December 2023. Since it still requires the creation and validation of test short masks prior to short detection, more up-front time and effort is needed than for ESD, and the masking of detected shorts is not performed automatically onboard. Nevertheless, for shorts that do not coincide with quadrant current elevation and can thus only be detected via OSD, this process is less onerous.

Perhaps even more significant are this capability's implications for short mask re-checking. Since the re-checking of previously masked lines to determine if the inciting shorts were transient (and could thus be unmasked to recover MSA operability) is essentially identical to the process for OSD but with subtractive rather than additive test masks, this mitigation for losses in MSA operability is increasingly feasible. See Section \ref{sec:future-development} for a description of plans for re-checking shorts.

\section{MICROSHUTTER ARRAY STATUS AND OPERABILITY}
\label{sec:msa-status-and-operability}

The MSA is used for a substantial portion of JWST's observing time: among the 17 JWST observing modes, MOS usage accounted for 14.8\% of accepted JWST Cycle 1 proposals\footnote{\url{https://www.stsci.edu/files/live/sites/www/files/home/jwst/science-planning/user-committees/jwst-users-committee/_documents/jstuc-0421-jwst-cycle1-review-package.pdf}} and 12\% of accepted JWST Cycle 2 proposals\footnote{\url{https://www.stsci.edu/contents/newsletters/2023-volume-40-issue-02/jwst-cycle-2-tac-results}} by time.\footnote{Long-range JWST science program processing is characterized by year-long proposal cycles, with Cycle 1 starting 1 July 2022. Note that proposals accepted for one cycle may in some cases be executed in a different cycle.}

\subsection{Magnet arm usage}
\label{sec:magnet-arm-usage}

\begin{figure} [htbp]
\begin{center}
\includegraphics[width=0.8\linewidth]{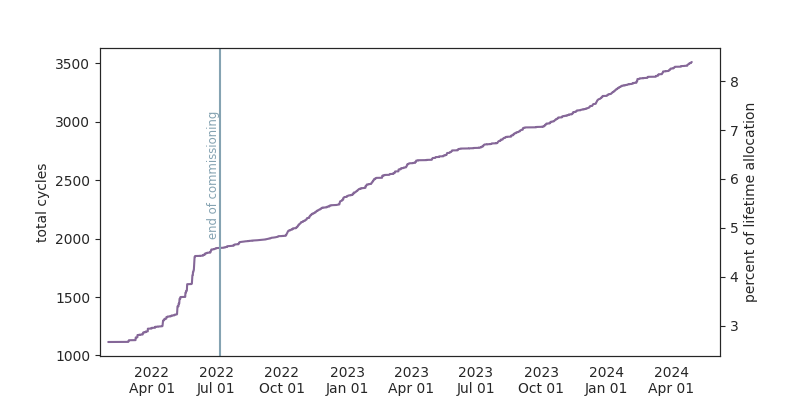}
\end{center}
\caption{Magnet arm usage in flight. One cycle is a full arm sweep from Primary Park to Secondary Park and back}
{ \label{fig:magarmusage} }
\end{figure} 

In Cycle 1 of normal operations, 806 MSA reconfigurations were performed over 365 days, for an average rate of 2.2 per day, which is one third the rate of reconfigurations during the 121 days of commissioning. In the 89\% of Cycle 2 that has elapsed at the time of writing, 771 reconfigurations have been performed, for an average rate of 2.4 per day. Relative to the irregular cadence of reconfigurations in commissioning and the early months of normal operations, the frequency of reconfigurations since then has been reasonably uniform.

Along with other components of NIRSpec including the filter and grating wheels and the calibration lamps, use of the MSA transport mechanism (magnet arm) is tracked as a ``Limited Life Item'' with lifetime allocations for each component in each mission phase established through qualification testing of an identical unit prior to integration at the instrument level. In the case of the magnet arm, the units of usage for the purpose of this tracking are cycles\footnote{Each reconfiguration is effected by one magnet arm cycle. Included in this count are reconfigurations to close all of the shutters at the end of any visit.}, where one cycle consists of round-trip travel of the magnet arm between Primary Park and Secondary Park positions. The total lifetime allocation for the magnet arm is 41852 cycles; as of 6 May 2024 it has completed 3527 cycles, or 8.4\% of its total lifetime allocation. Figure \ref{fig:magarmusage} shows the accumulation of magnet arm cycles since launch, with totals including usage during ground testing. Based on this usage rate of approximately 2\% of allocation expended per year of science operation, we do not expect the magnet arm to exceed its allocation of cycles within even the most optimistic projections of observatory operational lifetime.

\subsection{Mechanical operability status}
\label{sec:mechanical-operability-status}

A total\footnote{of the unvignetted field} of 20 shutters are currently identified as FO. A total of 41812 (18.5\%) are inoperable of which 24668 (10.5\%) are due to short masking (detailed in Section \ref{sec:electro-optical-operability-status}) and 17144 (7.6\%) are FC. Since the end of commissioning the number of FO decreased by 2 (see Table \ref{tab:fo}), and the FC increased by 1251, or 0.6\% (see Table ~\ref{tab:fc}).

\begin{table}
\caption{Failed Open (FO) shutters during the third Cryo-Vacuum test campaign (CV3), Optical Telescope element / Integrated Science instrument module (OTIS) test campaign, two epochs of commissioning (COMM), Cycle 1 (CYCLE 1), and the time of writing in Cycle 2 (CYCLE 2). For consistency, these numbers exclude vignetted FO shutters.}
\centering
\begin{tabular}{lrccccc}
\\
\textbf{FO} & & \textbf{Q1} & \textbf{Q2} & \textbf{Q3} & \textbf{Q4} & \textbf{Total} \\
\hline
Final CV3 & 2015 & 5 & 3 & 9 & 2 & 19 \\
Final OTIS & 2017 & 4 & 3 & 10 & 1 & 18 \\
Initial COMM & 2022 & 6 & 3 & 10 & 1 & 20 \\
Final COMM & 2022 & 6 & 3 & 12 & 1 & 22 \\
Final CYCLE 1 & 2023 & 5 & 3 & 11 & 1 & 20 \\
Time of writing CYCLE 2 & 2024 & 4 & 3 & 11 & 2 & 20 \\
\end{tabular}
\label{tab:fo}
\end{table}

\begin{table}
\caption{Failed Closed (FC) shutters during the third Cryo-Vacuum test campaign (CV3), Optical Telescope element / Integrated Science instrument module (OTIS) test campaign, two epochs of commissioning (COMM), Cycle 1 (CYCLE 1), and the time of writing in Cycle 2 (CYCLE 2). For consistency, these numbers exclude vignetted shutters. The CV3/OTIS counts differ from those presented in earlier analysis \cite{2018SPIE10698E..3QR}, which did not fully account for short masking. Note that the MSA shutter operability program suffered from a front end bug in Cycle 1 that prevented updating accounting of FC shutters.}
\centering
\begin{tabular}{$l^r^r^r^r^r^r}
\\
\multicolumn{1}{l}{\textbf{FC}} & \multicolumn{1}{c}{Q1} & \multicolumn{1}{c}{Q2} & \multicolumn{1}{c}{Q3} & \multicolumn{1}{c}{Q4} & \multicolumn{1}{c}{Total} & \multicolumn{1}{c}{\% FC} \\
\hline
CV3 & 1555 &  3215 &  6516 &  1892  &  13178   &  5.8\% \\
OTIS & 1607  & 3071 &  5999 &  2748  &  13425  &   5.9\% \\
Start COMM & 1605 &  3214 &  6027 &  4143 &   14989 & 6.6\% \\
End COMM & 1569 & 3328 & 5932 & 5064 & 15893 & 7.0\% \\
End CYCLE 1 & 1569  & 3328 & 5932  & 5064  & 15893   &   7.0\% \\
Time of writing (CYCLE 2) & 1582 & 3460 & 5945 & 6157 & 17144 & 7.6\% \\
\\
\\
\end{tabular}
\label{tab:fc}
\end{table}

\subsection{Electro-optical operability status}
\label{sec:electro-optical-operability-status}

\begin{figure} [htbp]
\begin{center}
\includegraphics[width=\textwidth]{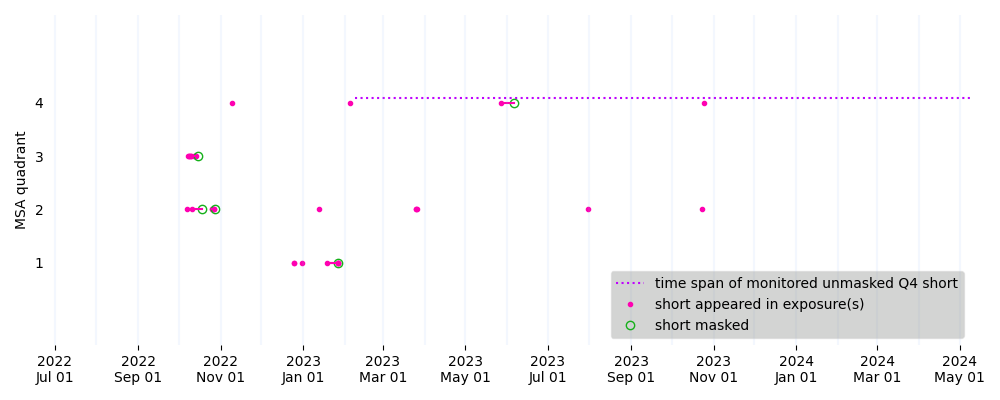}
\end{center}
\caption{Timeline of the appearances of shorts in normal operations, with horizontal lines connecting recurrences of what is judged—based on approximate location within a quadrant—to be the same short. The dotted line beginning in February 2023 indicates the time range of over which a low-level, as-yet-unmasked short in Q4 has recurred.}
{ \label{fig:shortstimeline}}
\end{figure}

As illustrated in Figure \ref{fig:shortstimeline}, the rate of appearances of shorts has generally decreased over the first two years of normal operations. A notable exception is the absence of shorts in the first three months following the end of commissioning, which may be partially explained by the relatively infrequent use of the MOS observing mode in that period.

Since commissioning, 26 visits have been approved by the JWST telescope time review board for repetition based on contamination from shorts experienced in the original execution. Note that in some cases the PI of a program with exposures impacted by shorts chooses not to request repetition of an affected visit because the program's goals could be accomplished without the exposures that were contaminated.

The majority of shorts that have produced contaminating glow in normal operations appear to have been transient. This transience and the fact that many of the shorts have occurred in clusters within short time spans suggests the presence of one or more floating particles alighting on the array briefly and temporarily creating a short before floating away—thereby dispelling the short—and then landing elsewhere on the array, inducing a short in a different location, perhaps in a repeating cycle. Along with a more concrete understanding of the transient nature of most shorts, since commissioning we have developed a better appreciation of the potential impacts on observatory schedule efficiency of activities of indeterminate duration like ESD. Some aspects of the initial concept of operations for mitigating shorts have evolved accordingly: 
\begin{itemize}
\item The urgency of mitigating shorts has decreased, particularly for shorts that are suspected to be transient based on their disappearance during an exposure instead of persisting until the next shutter reconfiguration. As mentioned in Section \ref{sec:background}, ESD is now executed as part of an OP, and early in Cycle 1 the appearance of a short customarily prompted the immediate development and uplink of a new ``intercept'' OP as quickly as possible to perform ESD and in some cases remove MOS visits. Given the newly-understood substantial probability of a new short turning out to be transient, this disruptive level of effort is no longer justified, particularly now that a flight software update (see Section \ref{sec:hold-mask-application}) prevents the appearance of shorts in IFS exposures.
\item Early in normal operations, ESD was often performed on all MSA quadrants rather than only the one(s) in which a short appeared in the course of previous observations. Not only would this practice potentially mask transient shorts that would not have resulted in glow in any science exposures, but it also exacerbates the intrinsic scheduling indeterminacy introduced by ESD activities. Now ESD is only performed on the quadrant in which a short has been observed.
\item It was initially conceived that ``preventative'' ESD could be added to an OP prior to particularly long or critical MOS visits. This is no longer part of our operations concept for all the reasons described above, including implicitly requiring all four quadrants to be tested.
\item The transience of many shorts suggests that the prospective re-checking of short masked lines could be quite likely to find lines that could be unmasked for the sake of increased operability.
\end{itemize}

\begin{figure} [htbp]
\begin{center}
\includegraphics[width=0.6\linewidth]{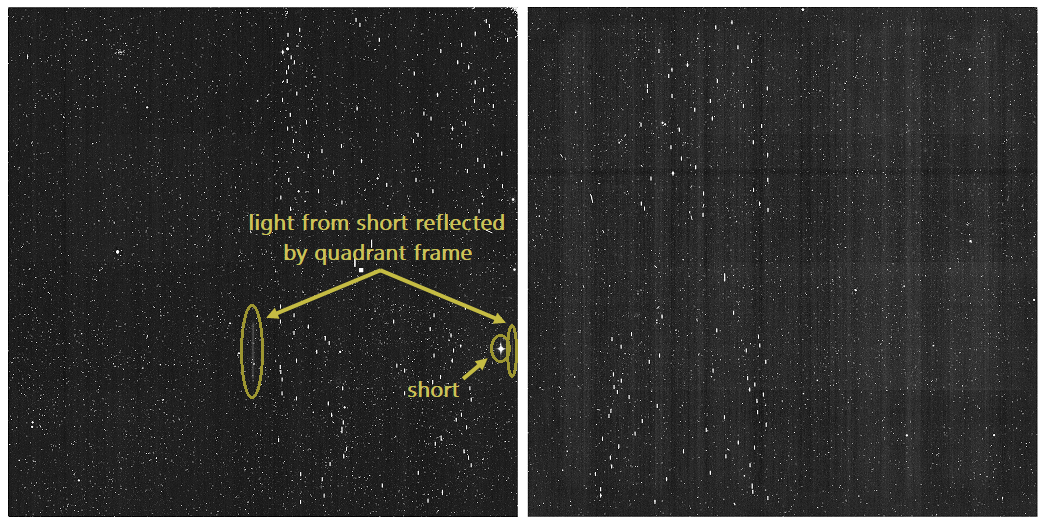}
\end{center}
\caption{Example of a target acquisition confirmation exposure with minimal contamination from a weak short}
{ \label{fig:weak-short-example} }
\end{figure}

\begin{table}
\caption{Timeline of short detection during the third Cryo-Vacuum test campaign (CV3), Optical Telescope element / Integrated Science instrument module (OTIS) test campaign, commissioning (COMM), and normal operations Cycles 1 and 2 (CYCLE 1 and CYCLE 2) thus far showing the number of rows/columns each episode added to the mask.}
\centering
\begin{tabular}{$l^l^r^c}
\\
\multicolumn{1}{l}{\textbf{Date}} & \multicolumn{1}{l}{\textbf{Description}} &
\multicolumn{2}{c}{\textbf{Added rows/columns}} \\
\hline
\multicolumn{4}{l}{\textbf{CV3}} \\
23 Nov 2015 & Electrical \#1 & 6 & \\
30 Nov 2015 & Electrical \#2 & 5 & \\
30 Nov 2015 & Optical \#1 & 2 & \\
09 Dec 2015 & Electrical \#3 & 2 & \\
03 Jan 2015 & Electrical \#4 & 2 & \\
06 Jan 2015 & Optical \#2 & 1 & \\
\multicolumn{2}{l}{\textbf{Total CV3}} & & \textbf{18} \\
\hline
\multicolumn{4}{l}{\textbf{OTIS}} \\
13 Aug 2017 & Electrical \#1 & 3 & \\
21 Aug 2017 & Optical \#1 & 1 & \\
02 Sep 2017 & Electrical \#2 & 3 & \\
09 Sep 2017 & Electrical \#3 & 0 & \\
11 Sep 2017 & Optical \#2 & 1 & \\
17 Sep 2017 & Optical \#3 & 1 & \\
\multicolumn{2}{l}{\textbf{Total OTIS}} & & \textbf{9} \\
\hline
\multicolumn{4}{l}{\textbf{COMM}} \\
27 Feb 2022 & Electrical \#1 & 2 & \\
09 Mar 2022 & Electrical \#2 & 1 & \\
23 Mar 2022 & Electrical \#3 & 2 & \\
26 Mar 2022 & Optical \#1 & 5 & \\
14 May 2022 & Optical \#2 & 1 & \\
20 May 2022 & Electrical \#4 & 1 & \\
02 Jun 2022 & Optical \#3 & 2 & \\
22 Jun 2022 & Electrical \#5 & 1 & \\
23 Jun 2022 & Electrical \#6 & 0 & \\
\multicolumn{2}{l}{\textbf{Total COMM}} & & \textbf{15} \\
\hline
\multicolumn{4}{l}{\textbf{CYCLE 1}} \\
15 Oct 2022 & Electrical \#1 & 1 & \\
27 Oct 2022 & Electrical \#2 & 1 & \\
11 Nov 2022 & Electrical \#3 & 0 & \\
01 Jan 2023 & Electrical \#4 & 0 & \\
26 Jan 2023 & Optical \#1 & 2 & \\
08 Feb 2023 & Electrical \#5 & 0 & \\
06 Jun 2023 & Electrical \#6 & 2 & \\
\multicolumn{2}{l}{\textbf{Total CYCLE 1}} & & \textbf{6} \\
\hline
\multicolumn{4}{l}{\textbf{CYCLE 2}} \\
04 Aug 2023 & Electrical \#1 & 0 & \\
30 Oct 2023 & Electrical \#2 & 0 & \\
\multicolumn{2}{l}{\textbf{Total CYCLE 2}} & & \textbf{0} \\
\end{tabular}
\label{tab:sds}
\end{table}

\begin{table}
\caption{Detailed statistics for the initial and final short masks of CV3, OTIS, commissioning (COMM), and Cycles 1 and 2 of normal operations thus far. Tabulated values include both the total number of masked rows/columns per quadrant (Q1–Q4, Total) and the number and percent of unvignetted shutters removed by the mask. CV3–OTIS: re-examination of the data led to two lines being removed from the mask. OTIS–COMM: the short mask remained unaltered and the slight decrease in shutters here arises from the change in identified vignetted shutters.}
\centering
\begin{tabular}{l r r r r r r r}
\\
 & \multicolumn{5}{c}{\textbf{Masked rows/columns}} & \multicolumn{2}{c}{\textbf{Masked shutters}} \\
 & \multicolumn{1}{c}{Q1} & \multicolumn{1}{c}{Q2} & \multicolumn{1}{c}{Q3} & \multicolumn{1}{c}{Q4} & \multicolumn{1}{c}{Total} & \multicolumn{1}{c}{\#} & \multicolumn{1}{c}{\%} \\
\hline
Start of CV3 & 20 & 14 & 10 & 25 & 69 & 14342 & 6.3\% \\
End of CV3 & 29 & 16 & 13 & 29 & 87 & 18431 & 8.2\% \\
Start of OTIS & 27 & 16 & 13 & 29 & 85 & 18123 & 8.0\% \\
End of OTIS & 29 & 21 & 15 & 29 & 94 & 20450 & 9.1\% \\
Start of COMM & 29 & 21 & 15 & 29 & 94 & 20440 & 9.1\% \\
End of COMM / Start of CYCLE 1 & 41 & 22 & 17 & 29 & 109 & 23628 & 10.5\% \\
End of CYCLE 1 / Start of CYCLE 2 & 43 & 24 & 18 & 31 & 116 & 24668 & 10.9\% \\
Time of writing & 43 & 24 & 18 & 31 & 116 & 24668 & 10.9\%  \\
\end{tabular}
\label{tab:sm}
\end{table}

One idiosyncratic case of a short that has surfaced in normal operations is a weak short in Quadrant 4 (see Figure \ref{fig:weak-short-example}) that appeared in February 2023 and has recurred often since. It did not produce detectable electrical current signatures. Because of its low impact on exposures relative to the potential decrease in operability that would be posed by masking it, this short has not been masked, and is instead being monitored for any changes in severity.

\subsubsection{Cross-instrument impacts}

\begin{figure} [htbp]
\begin{center}
\includegraphics[width=0.6\linewidth]{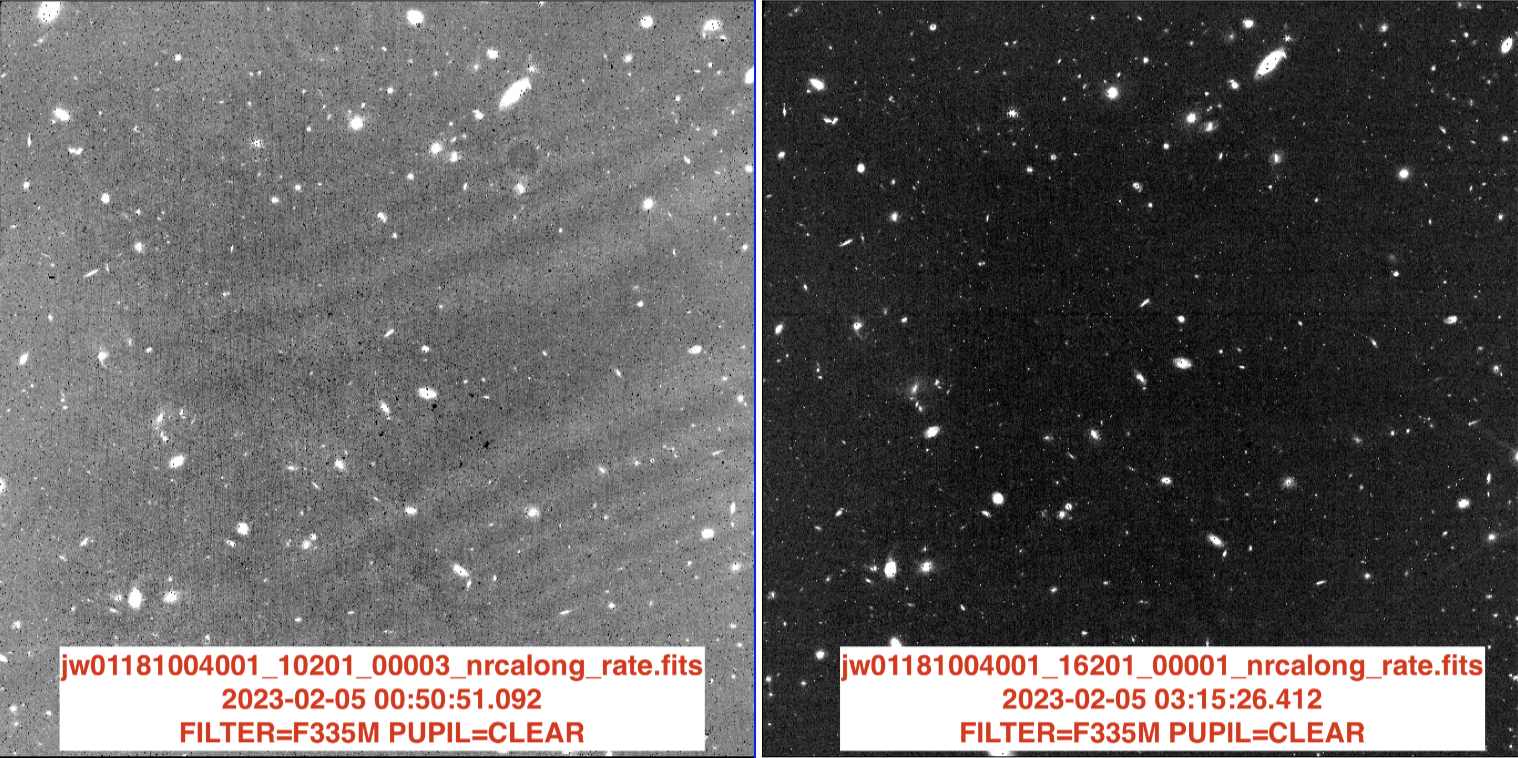}
\end{center}
\caption{Two NIRCam exposures in the same observation, sharing the same target and configuration. The exposure on the left was taken while an MSA short was contaminating NIRSpec exposures; the one on the right, taken several hours later, was not.}
{ \label{fig:nircam-contamination} }
\end{figure}

Intriguingly, in some cases glow from MSA shorts appears to have weakly contaminated NIRCam's exposures when the two instruments were exposing in parallel. NIRSpec MOS with NIRCam Imaging is a particularly popular coordinated parallel combination, and indeed some NIRCam exposures appeared to exhibit contamination (illustrated in Figure \ref{fig:nircam-contamination}) at the same time that a short on the MSA was producing significant infrared glow in a parallel instrument configuration. It is theorized that the telescope's secondary mirror can reflect light from an MSA short to the NIRCam light path, particularly onto NIRCam's long wavelength detectors and when NIRSpec's clear filter is in its own optical path. It may be possible to model and subtract the effects of this light in the NIRCam exposures, but since the accompanying NIRSpec exposures would at any rate be unusably contaminated and the visit likely to be repeated, developing these measures is not planned.

\subsection{Overall shutter operability status}
\label{sec:overall-shutter-operability-status}

\begin{figure} [htbp]
\begin{center}
\includegraphics[width=\linewidth]{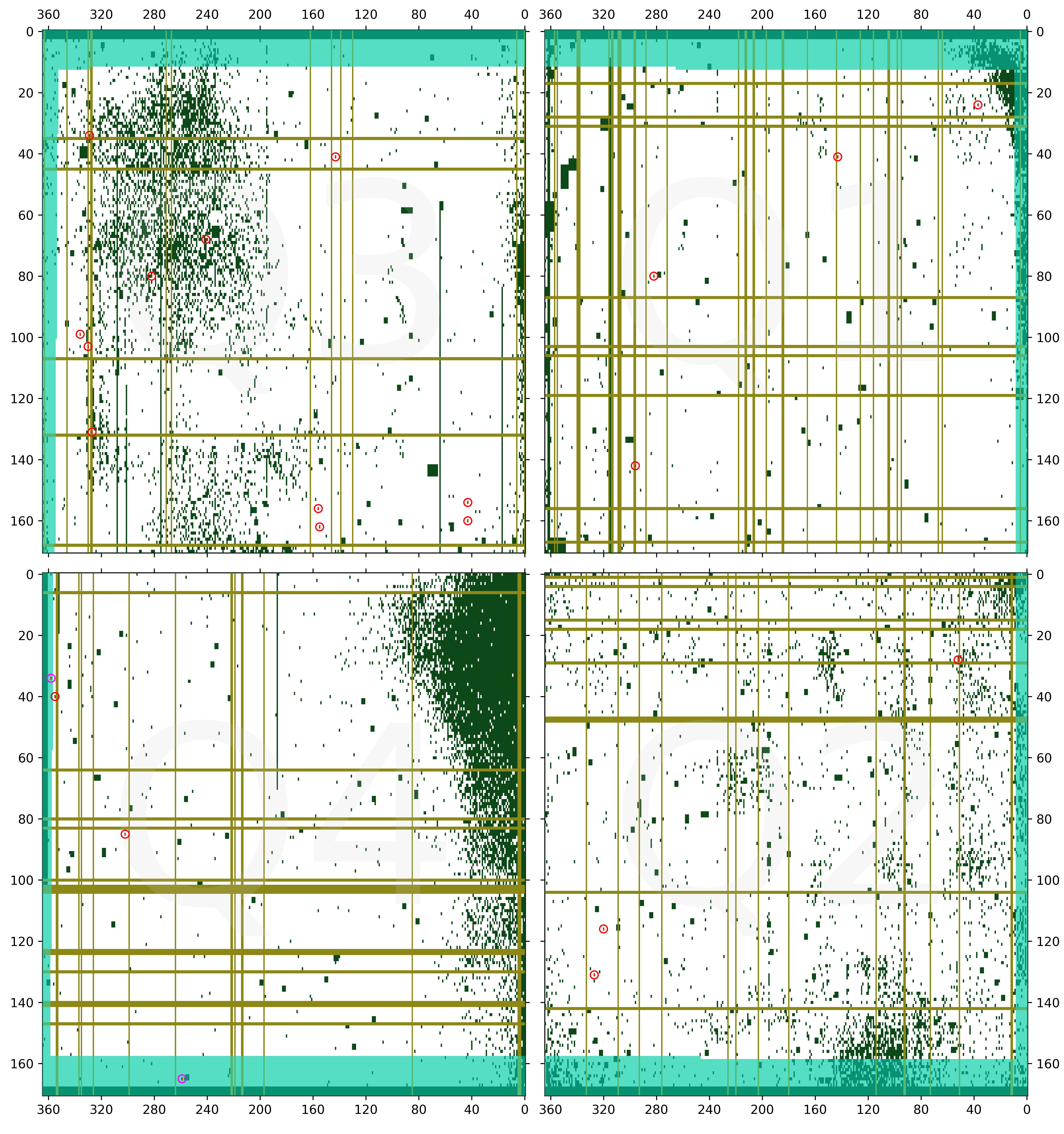}
\end{center}
\caption{Map of all inoperable shutters at the time of writing. Red circles identify failed open shutters. Dark green denotes failed closed shutters, in some cases due to plugs installed during component-level testing. Olive green denotes lines of shutters masked in short masks. Aqua colored areas are sections of the array vignetted by the field stop, rendered translucent to show failed and short-masked shutters behind the field stop.}
{ \label{fig:shuttermap}}
\end{figure}

As projected in pre-launch assessments \cite{2018SPIE10698E..3QR}, the increase in inoperable shutters since ground testing and commissioning has proven modest. At the time of writing, the latest shutter operability map update was made on 16 February 2024; the current map is portrayed in Figure \ref{fig:shuttermap}. Together with shutters included in short masks, the total number of inoperable shutters increased by 2291 or 1.2\% of all usable shutters since the beginning of normal operations. While it is true that an increasing number of inoperable shutters could negatively impact MSATA or the multiplexing capabilities of NIRSpec \cite{jakobsen2024onorbit}, we do not observe any significant impact at current operability levels. Even in the most empty regions of the sky, sufficient reference targets (stars or compact galaxies) can be identified for successful MSATA, and no observations using MSATA have been infeasible to schedule.

\begin{figure} [htbp]
\begin{center}
\includegraphics[width=0.8\linewidth]{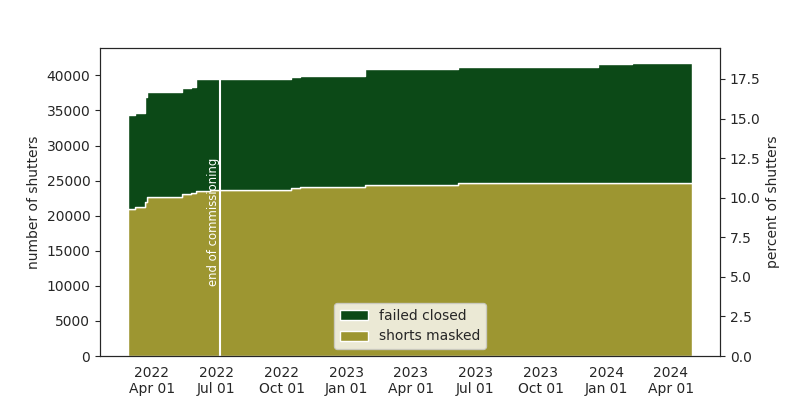}
\end{center}
\caption{Unusable shutter population in flight. These figures do not include vignetted shutters. The count of shorts-masked shutters includes shorts-masked shutters that are also failed closed.}
{ \label{fig:operabilitytrend}}
\end{figure}

\section{Fault Management}
\label{sec:fm}

JWST employs a layered fault management approach. For this reason, a variety of fault management actions could result in ``safe'' or powered-off states for the MSS. Observatory or ISIM-level fault management can power off all of ISIM, including the MSS, while instrument-level fault management may result in the MSS autonomously transitioning to a ``safe'' mode in the event of a command error, for instance, or in the powering off of the MSS in the event of a component issue. Each of these layers of fault management is software-driven. Hardware fault protection is also engineered into the MSS, allowing the flight software not just to power off one or more of the MSA quadrants in the event of an excessive current condition, but also to limit the current to the array itself via the use of resistors.

The ``safe'' mode of operation, also referred to as ``preferred safe'' configuration, for the MSS is a state in which the MCE remains powered, but power to the critical components of the MSS (i.e. the MSA quadrants and the magnet arm motor) is removed, and the shutters are closed.

Transitions of the MSS to this safe mode have occurred at various times during normal operations, usually as a result of OSS recognizing and responding to an error condition/exception, which can be the result of a commanding issue with any of the science instruments. Such a response includes transitioning most of the instruments into safe modes, though it is possible to transition just the MSS as a result of a NIRSpec-specific error condition.

\section{Resolved Issues}
\label{sec:resolved-issues}

In addition to passing a magnet across the shutters to program custom aperture masks, the micro-shutter assembly also serves to uncover the IFU aperture for IFS exposures and cover it when not in use by means of a semicircular tab mounted on the magnet arm. As design and testing of the MSA was primarily focused on micro-shutter operation, some impacts of this MSA-oriented paradigm on IFU operation have been uncovered in flight, as described in the following subsections.

\subsection{Electrical configuration after IFU aperture closure}
\label{sec:end-of-ifu-visit-hold-mask}

According to the logic implemented in the MSS hardware, when the magnet arm moves in the latching direction, the shutter open voltages are applied to the MSA quadrants, and no further electrical reconfiguration takes place until the subsequent programming-direction move is initiated. This behavior correctly allows for the opening of shutters when the magnet arm is swept over the array from the Primary Park position to the Secondary Park position, when the expectation is that it will be followed by a move in the programming direction, during which program voltages are applied and after which hold voltages are finally applied. The hardware-based logic does not distinguish this usage from moves in the latching direction that do not involve shutter reconfiguration, namely the move from the IFU position to the Primary Park position executed at the end of most IFS visits\footnote{If designated by the observer, an IFS visit can end with a pointing verification image through the MSA (in lieu of corrective target acquisition at the beginning of the visit) and so can involve full magnet arm sweeps over the shutters.} to cover the IFU aperture when IFS visits are not taking place. As a result, such moves leave open voltages applied to the MSA quadrants until the next NIRSpec visit (or commanding) that moves the magnet arm, which may be hours or days later. As NIRSpec's MSA is the first operational space-based micro-shutter array, the long-term effects of leaving non-quiescent voltages applied to the shutters are not well characterized, but with the understanding that the hardware is safest in its most electrically quiescent state we have conservatively chosen mitigation.

\begin{figure} [htbp]
\begin{center}
\begin{tabular}{c}
\includegraphics[width=0.75\linewidth]{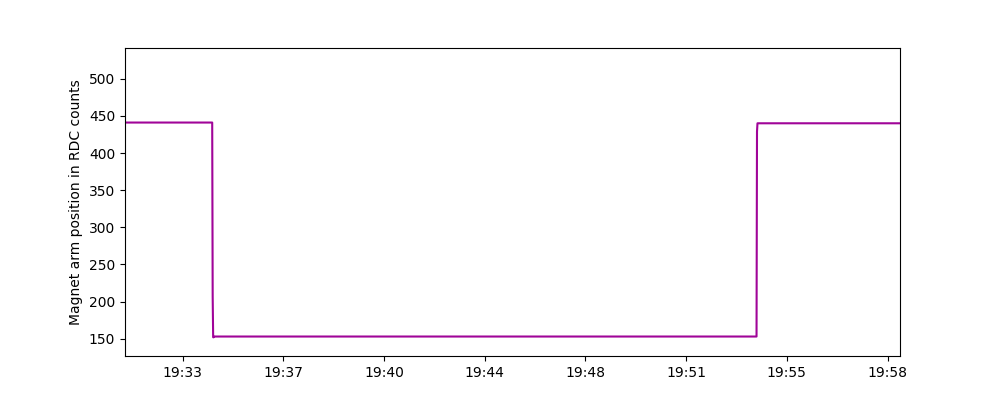} \\
\includegraphics[width=0.75\linewidth]{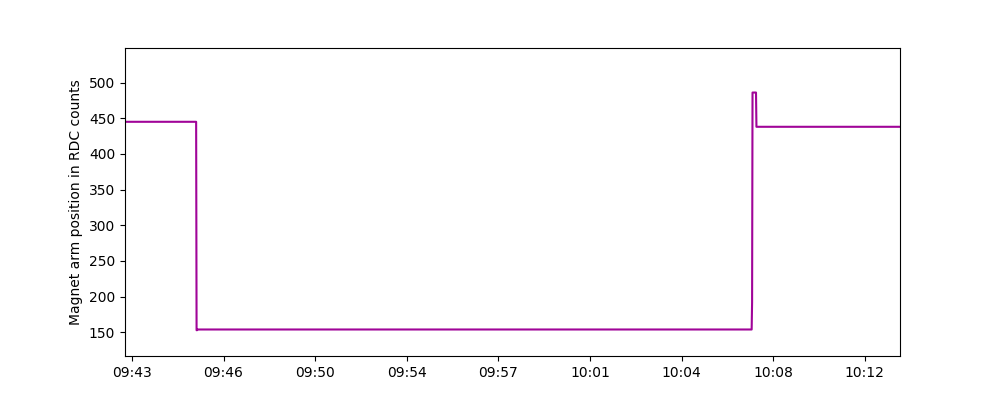}
\end{tabular}
\end{center}
\caption{Top: magnet arm position in the course of an IFS visit before introduction of the PPO position. Bottom: magnet arm position in the course of an IFS visit after introduction of the PPO position. In both cases the magnet arm starts at the PP position before uncovering the IFU aperture by moving to the IFU position. Before introduction of the PPO position, such an IFS visit would end with the magnet arm returning directly to the PP position, thereby leaving the IFU aperture covered. After the update, this return to the PP position comprises first a move from the IFU position to the PPO position and only then a move returning the magnet arm to the PP position, importantly accomplished with movement in the program direction.}
{ \label{fig:ppo_magarm_plot} }
\end{figure} 

Once the unintentional extended application of shutter open voltages after IFU visits was identified as undesirable not long after the start of Cycle 1, mitigation evolved in stages. Initially, real-time commanding of full magnet arm sweeps was used to reduce quadrant currents following IFS visits that did not end with a full magnet arm sweep. It would be an understatement to describe as ``logistically challenging'' the coordination of such commanding in DSN contacts as soon as possible following such IFS visits, at times when no NIRSpec observation was taking place or potentially could take place—which was non-trivial given that NIRSpec is the most used instrument on the observatory—and taking into account the potential for the actual execution of visits to occur at different times than expected according to the nominal OP timeline. Thus, in another example of the transition of some on-board activities from real-time commanding to OP-driven, this form of mitigation was replaced as quickly as possible with less onerous OP-initiated magnet arm sweeps, added manually by schedulers to each OP as supplemental five-minute internal visits. Still, this improved mitigation increased significantly the rate of limited-life magnet arm cycle usage and reduced observatory efficiency. As a long-term resolution, a change to OSS was implemented later in Cycle 1 to add a step to magnet arm moves from IFU position to Primary Park position: the magnetic arm is now moved from IFU position to the PPO position, which is located slightly less than two millimeters beyond the PP position, before being moved to the PP position as depicted in Figure \ref{fig:ppo_magarm_plot}. This last step moves the magnet arm in the programming direction to induce the MSS hardware to apply the hold voltage at the end of the visit. This update allows for the array to be placed into its most electrically quiescent state, thereby minimizing any risk of latent shorts appearing as a result of such moves.

\subsection{Application of hold mask after moves in programming direction}
\label{sec:hold-mask-application}

As mentioned in Section \ref{sec:msa-functionality}, the final step in programming a shutter pattern, after the magnet arm has returned to the Primary Park position, is the application of a ``hold mask'' that mitigates unwanted glow, including that from electrical shorts. The MSS flight software was designed to apply this mask at the end of any move from the Secondary Park position to any lower position below the array—i.e., after the micro-shutters have been programmed or released. It was not anticipated that this mask would be needed when the magnet arm is moved from the Primary Park position to the IFU position to uncover the IFU aperture, since this action does not release or program any shutters. However, as mentioned in the previous subsection, the voltage control logic in the MSS hardware does not distinguish between program-direction magnet arm movements based on the starting or destination positions; it indiscriminately applies a potential difference to the array during any program-direction movement, including this PP-to-IFU position case. It was found in commissioning that this potential difference can activate latent shorts on the shutter array, resulting in glow that can contaminate IFU exposures. In fact, 11 of the 26 visits approved for repetition due to short-based contamination were IFS visits.

Once this issue was recognized, identifying a solution was straightforward: the MSS flight software was patched to include the application of a hold mask after any program-direction move originating from the Primary Park Overshoot or Primary Park positions. Since this patch was deployed in June 2023, no IFS exposures have been contaminated by glow from MSA shorts.

\section{Open/Monitored Issues}
\label{sec:open-and-monitored-issues}

After two years of executing science observations, a few incompletely understood MSA behaviors have arisen. These are believed to be benign and have involved little to no mitigation effort. To better characterize these behaviors, they are included in daily monitoring by NIRSpec flight systems engineers and documented in JWST operations notes.

\subsection{Post-quadrant-power-on Quadrant 4 171-side RTN current elevation}
\label{sec:q4-171rtn-elevation-after-power-on}

Since onboard fault management powers off the micro-shutter assembly as part of every MSS transition to safe mode, restoring power is part of recovery to a nominal operating state. In every in-flight quadrant power-on, including all five in the course of safing event recoveries during commissioning (plus the initial in-flight power-on) and the fourteen so far in two years of normal operations safing recoveries, the initial Q4 171 RTN current has reached and sustained a level roughly twice the normal quiescent level of 1.5 mA.

A full sweep of the magnet arm over the array and back reliably returns this current to the quiescent level, so a step to perform such a move via real-time commanding is included in the MSS safing recovery procedure and has been added to the OSS safing recovery onboard script. There is no impact to normal operations as a result of this phenomenon, and its consistency suggests it is unrelated to the presence of MSA electrical shorts.

\subsection{Imprecise positioning after sequences of short moves}
\label{sec:pp-variability-after-short-moves}

As described in Section \ref{sec:end-of-ifu-visit-hold-mask}, short moves in the programming direction are employed to induce hold voltage and mask application below the micro-shutter array. These short moves—especially short in the case of moves from PPO to PP—often result in less precise and less repeatable final positioning of the magnet arm at the Primary Park position. While a situation in which the magnet arm were to stop over the array would be considered—as a potential risk to MSA hardware—an acute anomaly, the scale of deviations that seem to follow sequences of short moves since the introduction of the short PPO-to-PP move is far from positioning the magnet arm over the array.

However, the range of resting positions outside of which the MSS flight software considers a just-completed magnet arm move to have failed—at which point onboard fault management transitions the MSS to its safe mode and disables any science operations involving NIRSpec—was set during ground testing, before the PPO-to-PP move was introduced and when magnet arm resting positions exhibited less deviation. To avoid the unnecessary safing of the MSS and thus potential reduction of observatory efficiency caused by NIRSpec visits being needlessly skipped until the MSS is recovered and NIRSpec brought back online, the position tolerance window used by the MSS flight software is being expanded. This update will not reduce magnet arm resting position deviations, but it is expected to reduce the chance of an overly zealous fault management response to them.

\subsection{Uncommanded magnet arm movement}
\label{sec:uncommanded-magarm-movement}

As of this writing, three instances of changes in reported magnetic arm position (by one to four resolver counts) have been observed which are not associated with any commanding of the arm. In each case so far, the reported position increased—as opposed to decreased—and the reported position change occurred after fine guidance control had just been achieved for another instrument's visit and the appropriate data acquisition slot(s) assigned, and before the beginning of any instrument exposures. It is unclear whether the two instances of changes by a single resolver count reflect actual magnet arm movement or an idiosyncrasy of the resolver, but the four-count move corresponded with jitter in fine guidance imagery consistent with actual movement and was not attributable to any other cause. While positional changes of this magnitude are considered negligible and acceptable, having no discernible effect on normal operations, they are nonetheless of interest due to their ambiguous nature and are therefore flagged and tracked by flight systems engineers.

\section{Future Development}
\label{sec:future-development}

\subsection{Science-like operability assessment}

Ever since the mechanical shutter operability monitoring described in Section \ref{sec:monthly-operability-monitoring} was conceived, it has been understood that the use of checkerboard shutter patterns is a somewhat unreliable diagnostic for predicting the shutter failure rate in science-like configurations. Programming these $1\times1$ checkerboard patterns requires alternate voltage pulsing for every shutter along every row. In contrast with science-like configurations that command open typically a few hundred shutters, checkerboard patterns are maximally taxing on the MSA. Indeed, engineering telemetry corresponding to this checkerboard programming often includes quadrant currents on the 171 side that (momentarily) significantly exceed the highest levels reported in the programming of science patterns. 

Based on the tenuous value of $1\times1$ checkerboard patterns in operability monitoring given their demanding MSA programming and usage of four limited-life magnet arm cycles on a monthly basis, planning is underway to eliminate these patterns from the operability program. Instead, scripts would be developed to analyze science exposures taken in the course of normal operations to more realistically characterize the FC population.

Another aspect of FC operability open to more nuanced treatment is the stochastic element. Previous work includes an in-depth analysis of actual operability within commissioning-phase exposures, including a shutter-level distribution of success rates in response to open commands \cite{rawle22}. Although the vast majority of shutters in the analysis had a 100\% success rate, those that were not 100\% successful had a broad distribution of success rates, many of them higher than 80\%. It would in theory be possible to construct an operability map for use in MPT that accounts for the historically-derived probability of each particular shutter opening when commanded, rather than presenting a binary FC status. Observers could choose to use shutters with a substantial but not perfect record of success to add more targets to their exposures, with the understanding that these shutters have not always opened as commanded. Such a feature is only speculative at present but offers the possibility of more flexible pattern planning if desired and feasible later in the mission.

\subsection{Re-checking of short-masked lines}

As mentioned in Section \ref{sec:osd}, it is possible to test the persistence of masked shorts by creating test masks that differ from the current masks only in the removal of the lines that have masked the shorts being re-evaluated and using those test masks in a process otherwise identical to Optical Shorts Detection. The lines removed from the test masks used for those exposures not exhibiting glow from the short can be assumed to correspond to transient shorts that can be unmasked to recover MSA operability. This re-checking cannot be performed with Electrical Shorts Detection using the current onboard scripts because those scripts are incapable of using test mask files maintained in RAM by the instrument flight software. Still, since the introduction of the capability to apply test masks via the OP, re-checking short-masked rows and columns using the OSD method presents a practical opportunity to improve operability even when—as at present—operability is more than sufficient to support MSATA and the inclusion of observers' designated targets in their custom aperture masks. This re-checking is accommodated in a NIRSpec calibration program\footnote{Cycle 2: Program ID 4469, Cycle 3: Program ID 6646} that has not yet been instantiated but is expected to be used in at least a trial run in Cycle 3, once criteria for candidate lines to be re-checked are established.

\section{Summary}
\label{sec:summary}

Two years into the scientific operation of JWST, NIRSpec's MSA has performed at or above pre-launch expectations, delivering a multiplexing capability unique in space-based spectroscopy. Not only has it provided the basis for groundbreaking astronomical research, but as a pioneering use of MEMS technology, it serves as a potentially invaluable source of operational experience for the planning and design of future space-borne multi-object spectrographs. Indeed, managing micro-shutter operability and MSA electrical configuration issues is a complex effort with overlapping potential modes of failure, not to mention the kinds of behavioral idiosyncrasies that only emerge with months and years of normal science usage. We are pleased to report that the rate of new electrical shorts has declined dramatically since the beginning of normal operations; likewise, mechanical operability has proven satisfactory for observers' MOS programs. Avenues for optimizing operability remain to be explored if needed or desired, but based on the first two years of normal operations, we expect many more years of productive use of the NIRSpec MSA for trailblazing science.

\acknowledgments
NIRSpec was designed and built for ESA by Airbus Defence and Space GmbH in Ottobrunn, Germany, with the focal plane array and micro-shutter assembly provided by NASA Goddard Space Flight Center. We thank the dedicated engineers and scientists of ESA, Airbus, NASA, and STScI for their ongoing and enthusiastic support of NIRSpec operations. Bringing this work to the global astronomical instrumentation community would also not have been possible without the support of the JWST Mission Office. Lastly, we would like to acknowledge the JWST flight control team for conscientiously asking us to confirm a hard vacuum environment at L2 every time they are so prompted in the course of real-time MSA commanding.

\bibliography{references}
\bibliographystyle{spiebib} 

\end{document}